\documentstyle[epsfig,12pt,preprint,tighten,aps]{revtex}
\begin{document}

\draft

\begin{titlepage}
\rightline{June 2001}
\vskip 2cm
\centerline{\large \bf  
The mirror world interpretation of the 1908 Tunguska
event}
\vskip 0.3cm
\centerline{\large \bf  and other more recent events}
\vskip 1.1cm
\centerline{R. Foot\footnote{E-mail address:
foot@physics.unimelb.edu.au}}
\vskip .7cm
\centerline{{\it Research Centre for High Energy Physics}}
\centerline{{\it School of Physics}}
\centerline{{\it University of Melbourne}}
\centerline{{\it Parkville 3052 Australia}}
\vskip 2cm

\centerline{Abstract}
\vskip 1cm
\noindent
Mirror matter is predicted to exist if parity (i.e.
left-right symmetry) is a symmetry of nature. 
Remarkably mirror matter is capable of simply
explaining a large number of contemporary puzzles in 
astrophysics and particle physics including:
Explanation of the MACHO gravitational microlensing 
events, the existence of close-in extrasolar 
gas giant planets, apparently `isolated' planets, the solar, 
atmospheric and LSND neutrino anomalies, the orthopositronium 
lifetime anomaly and perhaps even gamma ray bursts. One 
fascinating possibility is that our solar system contains 
small mirror matter space bodies (asteroid or comet sized 
objects), which are too small to be revealed from their 
gravitational effects but nevertheless have explosive 
implications when they collide with the Earth. 
We examine the possibility that the 1908 Tunguska explosion in
Siberia was the result of the collision of a mirror
matter space body with the Earth. We point out that if 
this catastrophic event and many other similar smaller
events are manifestations of the mirror world then these 
impact sites should be a good place to start digging for 
mirror matter.  Mirror matter could potentially be 
extracted \& purified using a centrifuge and have many 
useful industrial applications.

\end{titlepage}
\noindent

One of the most natural candidates for a symmetry of nature is 
parity (i.e. left-right) symmetry.  While it is an established 
experimental fact that parity symmetry appears broken by the 
interactions of the known elementary particles, this however does not 
exclude the possible existence of exact unbroken parity symmetry in nature. 
This is because parity (and also time reversal) can be exactly conserved 
if a set of mirror particles exist\cite{ly,flv}.  The idea is 
that for each ordinary particle, such as the photon, electron, proton
and neutron, there is a corresponding mirror particle, of exactly 
the same mass as the ordinary particle. 
For example, the mirror 
proton and the ordinary proton have exactly the same mass.
Furthermore the mirror proton is stable for the same reason that 
the ordinary proton is stable, and that is, the interactions of 
the mirror particles conserve a mirror baryon number.
The mirror particles are not produced (significantly)
in Laboratory experiments 
just because they couple very weakly to the ordinary particles. 
In the modern language of gauge theories, the mirror particles 
are all singlets under the standard $G \equiv SU(3)\otimes SU(2)_L 
\otimes U(1)_Y$ gauge interactions. Instead the mirror
fermions interact with a set of mirror gauge particles,
so that the gauge symmetry of the theory is doubled,
i.e. $G \otimes G$ (the ordinary particles are, of 
course, singlets under the mirror gauge symmetry)\cite{flv}.
Parity is conserved because the mirror fermions experience
$V+A$ mirror weak interactions and the ordinary fermions 
experience the usual $V-A$ weak interactions.  Ordinary and mirror
particles interact with each other predominately by gravity only.

At the present time there is a large range of experimental evidence
supporting the existence of mirror matter (for a review
see Ref.\cite{puz}).  Mirror matter is necessarily stable and 
dark and appears to provide a viable candidate for the inferred
dark matter in the Universe\cite{blin}
as well as having important implications for early
Universe cosmology\cite{blin,eu}. 
Mirror dark matter also has self interactions
just like ordinary matter 
which may allow it to escape the fate of
collisionless cold dark matter candidates such as
hypothetical neutralinos which now appear to be ruled out by
the observations\cite{ruledout}.  Moreover, mirror matter,
like ordinary matter can form stars, planets and smaller
bodies and there is interesting evidence for all these things.
In particular mirror stars are a natural candidate\cite{ii} for
the observed MACHO gravitational microlensing events\cite{macho}.
Furthermore mirror planets would provide a simple explanation\cite{mp} for
the existence of close-in extrasolar planets which has been
puzzling astronomers since their unexpected discovery
in 1995\cite{queloz}. There is also evidence that
the `dynamical mirror image' system 
of an ordinary planet orbiting a mirror star has also been
observed but interpreted as an `isolated' planet because
light from the mirror star was not detected\cite{iso}.

The significance of mirror matter for astrophysics
and cosmology is clear, perhaps of equal importance though
is the implications of mirror matter for particle physics. 
While ordinary and mirror matter interacts with each other
predominately by gravity, small non-gravitational interactions
are actually possible.  Due to
constraints from gauge symmetry, renormalizability and parity
symmetry it turns out that there are only 3 ways in which
ordinary and mirror matter can interact with each other
(besides gravity)\cite{flv,flv2}. This is via photon - mirror photon
kinetic mixing, Higgs - mirror Higgs interactions and
via ordinary neutrino - mirror neutrino mass mixing
(if neutrinos have mass). While Higgs - mirror Higgs
interactions will be tested if or when the Higgs particle
is discovered, there is currently strong evidence
for photon - mirror photon kinetic mixing and also
ordinary neutrino - mirror neutrino mass mixing.

A simple consequence of the parity symmetry is that
each of the ordinary neutrinos ($\nu$) will oscillate maximally into
its mirror partner ($\nu'$)\cite{flv2,P,b2}. 
This provides a very elegant
explanation for the solar neutrino puzzle since
the maximal $\nu_e \to \nu'_e$ oscillations imply an approximate 
50\% flux reduction for a large range of
$\delta m^2$ which is in broad agreement with the solar 
neutrino data\cite{phen,recent}. Moreover this 
solution predicted the approximate energy independent 
recoil electron energy spectrum observed by super-Kamiokande\cite{sk}
as well as the $\sim \ 50\%$ flux reduction found in
the Gallium experiments\cite{gal}.
In the case of the atmospheric neutrino anomaly the
inferred 50\% reduction of up-going $\nu_\mu$ is also
nicely explained by maximal $\nu_\mu \to \nu'_\mu$ 
oscillations\cite{fvy}.  If the solar and atmospheric neutrino 
anomalies are due to oscillations into mirror neutrinos
then oscillations between generations can be governed
by small mixing angles which seems theoretically
most natural. This reasoning is supported by
the LSND experiment which has provided strong evidence
for small angle $\nu_e \to \nu_\mu$ oscillations\cite{lsnd}.  

It is true, though, that the solution to the neutrino physics
anomalies implied by the mirror matter theory does not give
a perfect fit to every neutrino experiment. 
However this is probably a good thing, since it is unlikely
that every experimental measurement is correct. 
In the case of solar neutrinos,
the low Homestake result (1/3 c.f. 1/2 in the 5 other 
solar neutrino experiments)
and also the recent SNO results\cite{sno} do not favour the
simplest mirror matter solution.
In addition the atmospheric data slightly prefer $\nu_\mu \to \nu_\tau$
to $\nu_\mu \to \nu'_{\mu}$\cite{skxx} (although the
extent to which $\nu_\mu \to \nu'_\mu$ is disfavoured depends
significantly on how the data is analysed\cite{foot2000}).
Because these disfavouring results are only at the 1.5-3.3 sigma
level (and are largely dominated by systematics) they do not
provide a strong case against the mirror matter theory.
Importantly things will eventually become clear as more
accurate measurements are done. The forthcoming $NC/CC$ SNO
measurement should provide a solid result one way or the other.

Another important way  that ordinary and mirror
matter can interact with each other is via photon -
mirror photon kinetic mixing.  In field theory
this is described by the interaction 
\begin{equation}
{\cal L} = {\epsilon \over 2}F^{\mu \nu} F'_{\mu \nu},
\label{ek}
\end{equation}
where $F^{\mu \nu}$ ($F'_{\mu \nu}$) is the field strength 
tensor for electromagnetism (mirror electromagnetism).
This type of Lagrangian term is gauge invariant 
and renormalizable and can exist at tree level\cite{fh,flv}
or maybe induced radiatively in models without $U(1)$ 
gauge symmetries (such as grand unified theories)\cite{bob,gl,cf}.
One effect of ordinary photon - mirror photon kinetic mixing
is to give the mirror charged particles a small electric
charge\cite{bob,gl,flv}. That is, they couple to ordinary photons with
electric charge $\epsilon e$.

The most important experimental constraint on photon - mirror
photon kinetic mixing is that it modifies the properties
of orthopositronium\cite{gl}. This effect arises due to
radiative off-diagonal contributions to the
orthopositronium, mirror orthopositronium
mass matrix. This means that orthopositronium oscillates
into its mirror partner. Decays of mirror
orthopositronium are not detected 
experimentally which effectively increases the observed 
decay rate\cite{gl}. Because collisions of orthopositronium destroy
the quantum coherence, this mirror world effect is most
important for experiments which are designed such that the
collision rate of the orthopositronium is low\cite{gn}.
The only accurate experiment sensitive to the mirror
world effect is the Ann Arbour vacuum cavity experiment\cite{vac}.
This experiment obtained a decay rate of 
$\Gamma_{oPs} = 7.0482 \pm 0.0016 \ \mu s^{-1}$.
Normalizing this measured value with the recent
theoretical value of $7.0399 \ \mu s^{-1}$ \cite{theory} gives
\begin{equation}
{\Gamma_{oPs}(exp) \over \Gamma_{oPs}(theory)} =
1.0012 \pm 0.00023
\end{equation}
which is a five sigma discrepancy with theory.
It suggests a value $\epsilon \simeq 10^{-6}$
for the photon - mirror photon kinetic mixing\cite{fg}.
Taken at face value this experiment is strong evidence
for the existence of mirror matter and hence
parity symmetry. It is ironic that the last time
something important was discovered
in high energy physics with a table top experiment was
in 1957 where it was demonstrated that the ordinary
particles by themselves appear to violate parity symmetry.

Of course this vacuum cavity experiment must be
carefully checked by another experiment to make
sure that mirror matter really exists. Actually
this is quite easy to do.  With the 
largest cavity used in the experiment of Ref.\cite{vac} 
the orthopositronium
typically collided with the cavity walls 3 times before
decaying. If the experiment was repeated with a larger cavity 
then the mirror world effect would be larger
because the decohering effect of collisions would be
reduced. For example if a cavity 3 times larger could be used
(which means that the orthopositronium would typically collide
with the walls just once before decaying) then
the mirror world would predict an effect 3 times larger. 

There are several important implications of
photon - mirror photon kinetic mixing with 
the relatively large value of $\epsilon \simeq 10^{-6}$
suggested by the orthopositronium vacuum experiment.
These include:
\begin{itemize}
\item
Exploding mirror stars (mirror supernova) will emit
a burst of (ordinary) gamma rays.
This would occur because at the temperatures 
$\sim 10 \ {\rm MeV}$ reached at the center of
a typical supernova explosion the kinetic mixing will 
convert $e'^+ e'^- \to  e^+ e^-$ which subsequently
produces a relativistic fireball, which seems to qualitatively
explain many of the features of the
observed gamma ray bursts\cite{blin2}.
\item
Such a large value of $\epsilon \approx 10^{-6}$ will lead
to the light mirror particles ($e'^{\pm}, \gamma', \nu'$)
being brought into equilibrium with the ordinary particles above
$T = 1 \ {\rm MeV}$ in the early Universe\cite{cg}.
While this is not a problem for the recent 
measurements\cite{boo} of the Cosmic Microwave Background\cite{steen}, 
it does suggest that standard BBN needs modification.
For example, there might exist a large electron neutrino
asymmetry which can compensate for the faster expansion
rate leading to acceptable values of the light element 
abundances\cite{xxx1}. Another possibility is 
that there might
exist a large negative cosmological constant
which will slow down the expansion rate at $T \sim 1 \ {\rm MeV}$
\cite{b2}.

\item
Mirror stars can become visible if they have some
embedded ordinary matter. This is because the ordinary
matter is heated by the mirror matter though photon -
mirror photon kinetic mixing. Maybe the recently observed
halo white dwarfs \cite{oppen} (which are controversial\cite{cont}) 
are really mirror stars\cite{fiv} or even mirror white dwarfs.
Because of their age they may have accreted enough ordinary
matter to be observable.

\end{itemize}

Perhaps the most remarkable possibility though is
that there is some significant amount of mirror matter in 
our solar system. 
We don't know enough about the formation of the 
solar system to be able to exclude the existence
of a large number of  Space Bodies (SB) made
of mirror matter if they are small
like comets and asteroids. The total mass of 
asteroids in the asteroid belt is estimated to be only
about 0.05\% of the mass of the Earth. A similar or
even greater number of mirror bodies, perhaps orbiting in a
different plane or even spherically distributed
like the Oort cloud
is a fascinating and potentially explosive possibility\footnote{
Large planetary sized bodies are also possible if they are 
in distant orbits\cite{silnem}.}
if they collide with the Earth.
The possibility that such collisions occur and may
be responsible for the 1908 Siberian explosion (Tunguska
event) has been speculated in Ref.\cite{puz}.
The purpose of this paper is to study this possibility
in detail and to point out the important ramifications
of this idea which is 
that mirror matter should be present in the ground
at the `impact' sites and could be extracted as we
will discuss.

If such small mirror bodies exist in our
solar system and happen to
collide with the Earth, what would be the consequences?
If the only force connecting mirror matter with
ordinary matter is gravity, then the consequences
would be minimal. The mirror SB would simply
pass through the Earth and nobody would know about it
unless it was so heavy as to gravitationally affect
the motion of the Earth. However if there is photon -
mirror photon kinetic mixing as suggested by the 
orthopositronium vacuum cavity experiment, then the 
mirror nuclei (with $Z'$ mirror protons) will effectively have 
a small ordinary electric charge $\epsilon Z' e$. This means that 
the nuclei of the mirror atoms of the SB will undergo 
Rutherford scattering off the nuclei of the atmospheric nitrogen 
and oxygen atoms. In addition ionizing interactions can
occur which can ionize both the mirror atoms of the space body 
and also the atmospheric atoms.
The net effect is that the kinetic energy of
the SB is transformed into light and heat (both ordinary
an mirror varieties) and a component
is also converted to the atmosphere in the form of a shockwave,
as the forward momentum of the SB is transferred
to the air which passes though or near the SB.

What happens to the mirror matter SB 
as it plummets towards the Earth's surface depends on
a number of factors such as its initial velocity, size, 
chemical composition and angle of trajectory.
Of course all these uncertainties
occur for an ordinary matter SB too.
Interestingly it turns out that for the value of the
kinetic mixing suggested by the Orthopositronium 
experiment, $\epsilon \approx 10^{-6}$, the air resistence
of a mirror SB 
in the atmosphere is roughly the same as an ordinary
SB assuming the same trajectory, velocity
mass, size and shape (and that it remains intact).  
This occurs because
the air molecules will lose their relative forward momentum
(with respect to the SB)
within the SB itself because of the Rutherford scattering
of the ordinary and mirror nuclei as we will show in a moment.
(Of course the atmospheric atoms still have random thermal motion). 
This will lead to a drag force
of roughly the same size as that on an ordinary matter SB,
implying an energy loss rate of
\begin{equation}
{dE \over dx} = C_d \rho_{air} A {v^2 \over 2} \ ,
\label{drag}
\end{equation}
where $\rho_{air}$ is the density of the air, $v$ is the 
velocity of the SB and $A$ is the cross sectional area.
The drag coefficient, $C_d$ is of order unity - its
precise value depending on the shape of the body.
We will take $C_d \sim 1$.  
Eq.(\ref{drag}) is a standard result and quite easy to
derive: The pressure of the atmosphere on the surface 
of the body increases linearly
with the velocity of the body. Also the number of atoms
striking the surface will increase linearly with the air density 
and also velocity (since the volume that the body sweeps 
out in a given time  $t$ is just $Avt$).
Eq.(\ref{drag}) implies
that the bodies velocity decreases exponentially
with distance ($x$),
\begin{equation}
v = v_i e^{-x/D}\ ,
\end{equation}
where $v_i$ is its initial velocity and
\begin{eqnarray}
D =  {2R\rho_{SB}\over C_d \bar \rho_{air}} 
\sim  10 \left( {R \over 5 \ {\rm meters}}\right)
\left( {\rho_{SB} \over 1 \ g/cm^3}\right)
\ {\rm km}.
\label{DD}
\end{eqnarray}
In this equation, $\rho_{SB}$ is the density of the 
SB and $R \equiv V/A$
is the `size' of the body ($V$ is its volume).
Note that we have used $\bar \rho_{air} \approx 10^{-3} \ g/cm^3$ which
is the air density at about 5 km altitude (the
density at sea level is about twice this value) for a rough
estimate of the mean density encountered as it travels through
the atmosphere.
The above calculation shows that the rate of energy
loss of the SB in the atmosphere depends on its
size and density.  
If we assume a density of $\rho_{SB} \simeq 1 \ g/cm^3$ which
is approximately valid for a mirror SB made of cometary material
(such as mirror ices of water, methane and/or ammonia)
then the body will lose most of its kinetic energy in the atmosphere
provided that it is less than roughly 5 meters in diameter.
Of course things are complicated because the the SB will undergo
mass loss (ablation) and also
potentially fragment into smaller 
pieces and of course potentially melt \& vaporize.
Thus even a very large body (e.g. $R \sim 100$ meters as estimated
for the Tunguska explosion) can lose its 
kinetic energy in the atmosphere if it fragments into small pieces.

An important difference between an ordinary and mirror SB
is the rate and way in which it fragments, heats up and undergoes
ablation because these
properties depend very much on the interactions between
the SB and the atmosphere. An ordinary matter
SB undergoes huge pressure on its surface when it enters the atmosphere with
cosmic velocity ($\sim \ 30 km/s$) while in the case of a
mirror matter body
the effects of the pressure are distributed within the body to some
extent, rather than just at the very surface. 
Let us now examine this in more detail.

Assume that the mirror matter SB is composed of
atoms of mass $M_{A'}$ and the air is composed of atoms of mass $M_{A}$.
The (mirror) electric charge in units of $e$ of the (mirror) nuclei, which
we roughly assume to be half neutrons and half protons,
will be $Z = M_A/{2M_P}$ ($Z' = M_{A'}/{2M_P}$),
where $M_P$ is the proton mass.
Let us assume that the trajectory of the SB
is a straight line along the $\hat{z}$ axis of our co-ordinate 
system.  In the rest frame of the SB,
the change in forward momentum of each of the on-coming 
atmospheric atoms is then\footnote{
The following equation is valid provided that
$M_{A'} \gg M_{A}$ but our conclusions will remain roughly the
same for other cases of interest such as
for $M_{A'} \sim M_{A}$.}
\begin{equation}
{dP_z \over dt} = \Gamma_{coll} M_A (v\cos\theta - v) = 
-2\Gamma_{coll} M_A v \sin^2 {\theta \over 2}\ ,
\label{essendon}
\end{equation}
where $\theta$ is the scattering angle in the
rest frame of the SB
and $\Gamma_{coll}$ is the collision rate of the
atmospheric atom with the mirror atoms in the SB. 
Of course the collisions also generate transverse momentum
(i.e. in the $\hat{x},\ \hat{y}$ directions) which
is reduced by thermalization effects
as the atoms in the atmosphere interact with themselves.
For the present calculation we are only interested in 
the relative net momentum between the SB and the
atmosphere and we can neglect this transverse motion in
a rough approximation (which means that we
can replace $v$ by $v_z$ below).  The collision rate
$\Gamma_{coll}$ is given in terms of the
cross section, relative velocity and number density in the
usual way:
\begin{equation}
\Gamma_{coll} = \sigma v_z \left( {\rho_{SB} \over M_{A'}}\right).
\end{equation}
Thus Eq.(\ref{essendon}) becomes
\begin{equation}
{dP_z \over dt} = -2\left({M_A \over M_{A'}}\right)
\int {d\sigma \over d\Omega} \rho_{SB} v_z^2 \sin^2
{\theta \over 2} \ d\Omega\ .
\label{dp}
\end{equation}
There are various different processes which can contribute to
the scattering cross section. For the velocities of 
interest, $v \stackrel{<}{\sim} 70\ km/s$,
the cross section is dominated by Rutherford scattering\footnote{
Although the cross section is dominated by Rutherford scattering,
ionizing collisions may also be important for generating light and
perhaps may also allow the body to build up electric 
charge within\cite{cep}.}
of the mirror nuclei of effective electric charge $\epsilon Z' e$
off the ordinary nuclei of electric charge $Ze$ modified
for small angle scattering by the screening effects
of the atomic electrons (at roughly the 
Bohr radius $r_0 \approx 10^{-8}\ {\rm cm}$). It is  
given by (see e.g. \cite{merz})\footnote{
We use standard particle physics units $h/(2\pi) = c = 1$ unless
otherwise stated. }:
\begin{equation}
{d\sigma \over d\Omega} = 
{4M^2_A \epsilon^2 e^4 Z^2 Z'^2 \over 
(4M_A^2 v^2_z \sin^2 {\theta\over 2} + 1/r_0^2)^2}.
\label{ds}
\end{equation}
Thus we obtain from Eq.(\ref{dp}) and Eq.(\ref{ds})
the following differential equation for 
the distance travelled by each atmospheric atom
($z$) within the SB :
\begin{equation}
{dP_z \over dt} = M_{A} \ v_z {dv_z \over dz}
\sim Z^2 Z'^2\rho_{SB} 
{\epsilon^2 e^4 4\pi \over M_{A'}M_A v_z^2} log_e \left({1 \over
M_A v_z r_0}\right),
\label{11}
\end{equation}
which is valid for $M_A v r_0 \gg 1$. For $M_A \approx 15M_P$,
$M_A v r_0 \approx 50 (v/30\ km/s)$ which means
that the above equation is approximately valid
for the velocities of interest (the initial 
velocity, $v_i$, of a SB is typically 
between 15 and 60 km/s).
Solving the above differential equation (neglecting
the log factor which is of order 1) we find that the
relative motion between the air molecules and SB 
is lost (upto random thermal motion) after travelling a distance
within the SB of
\begin{equation}
z \sim {v^4 M^2_{A}M_{A'} \over 16\pi Z^2 Z'^2 \rho_{SB} \epsilon^2 e^4}
\sim \left({10^{-6} \over \epsilon}\right)^2 \left({v \over 30 
\ km/s}\right)^4 \  \ {\rm centimeters}, 
\label{DDD}
\end{equation}
where we assumed $\rho_{SB} \approx 1 \ g/cm^3$
and $M_A \approx M_{A'} \approx 15M_P$ (with $Z \approx Z' \approx 7$).
For $\epsilon = 10^{-6}$, Eq.(\ref{DDD}) indicates
that the atmospheric atoms lose essentially all of their
relative momentum (of course they still have thermal motion)
after penetrating a distance of the order of a few centimeters 
into the SB. (This distance may be somewhat greater for
a body made of a heavy element such as mirror iron). If 
the SB remains intact then the above
result implies that the air resistence 
of the mirror SB through the atmosphere is
roughly the same as that of an ordinary matter
SB, as we already assumed earlier and have now proved. 
This {\it does not} mean that only the outer regions of the mirror SB
will be heated by the atmosphere. The atmospheric atoms
still have rapid thermal motion which will penetrate
deep into the mirror SB. This is of course
completely unlike a SB made of ordinary matter which
remains cool inside. This `internal
heating' of the mirror SB should make it easier for
the body to fragment and/or possibly build up
enough internal pressure to explode. 
However, because the huge pressure from
the atmosphere is dissipated over some distance within
the body rather than just at its surface, the rate of
ablation of a mirror SB may be significantly less
than that of an ordinary SB. 

Incidentally, if $\epsilon \stackrel{<}{\sim} 10^{-8}$
instead of the value $10^{-6}$ indicated by the orthopositronium
vacuum cavity experiment, a small or moderate sized
SB would not lose significant energy in the
atmosphere because the atmospheric atoms
would pass through the body without losing much of their
relative momentum. In this case the SB would
release most of its energy underground in the Earth's crust.
The distance over which this would occur
would simply be given roughly by Eq.(\ref{DDD})
with the replacement $\rho_{SB} \to \rho_E$
($\rho_E$ is the density of the Earth)
and $M_A \leftrightarrow M_{A'}$, which is
\begin{eqnarray}
L \sim   {v^4_i M_{A'}^2 M_{A} \over 16\pi \rho_E Z^2 Z'^2 
\epsilon^2 e^4 } 
\sim  \left({v_i \over 30 \ km/s}\right)^4 \left({10^{-9} 
\over \epsilon}\right)^2 \ km,
\end{eqnarray}
which was advertized earlier in Ref.\cite{puz}.

Returning to the most interesting case of large photon
- mirror photon kinetic mixing, $\epsilon \simeq 10^{-6}$ which
is indicated by the orthopositronium experiment, our earlier
calculation suggests that most of the kinetic energy of 
a mirror matter SB is released in the atmosphere like
an ordinary matter SB if it is not too big ($\stackrel{<}{\sim}
5$ meters) or fragments into small objects.
It seems to be an interesting candidate to explain the 1908
Tunguska explosion (as well as smaller similar events as
we will discuss in a moment).  The Tunguska 
explosion toppled approximately 2,100 square kilometers 
of trees in a radial pattern (i.e. like
spokes on a wheel) with an atmospheric release of 
energy estimated to be the TNT equivalent of roughly 1000 
atomic bombs\cite{tung}.  There was also evidence that the 
inner 300 square kilometers of trees was burned from 
above.  The broad features of the event suggest a huge
explosion in the atmosphere at an altitude of between about 
2.5 and 9 km which produced a downward going spherical 
shockwave\cite{tung}.
The spherical shockwave toppled the trees in the radial pattern
and the heat from the explosion caused the flash burn
of the trees\cite{tung}. 
An interesting feature of this event is the 
lack of any extraterrestrial fragments
or any (ordinary) crater(s). The estimated
mass of the SB is of the order of 100 thousand 
tons\cite{tung}.  That is no typo. 
It is a remarkable result that such a large amount of 
extraterrestrial material
apparently vanished without leaving behind significant remnants.
Over the last 75 years about 35 scientific
expeditions to the Tunguska site have been made with many
types of search techniques, but all coming back
empty handed. There have also been searches
for microparticles 
in tree resin with some success\cite{rc}.
However their tiny abundance 
is hardly consistent with what might have been 
expected.
It seems therefore to be a real possibility that the Tunguska 
event was due to a mirror matter SB
which would not leave any ordinary fragments (the
observed microparticles, if there are indeed of extraterrestrial origin,
may simply be due to 
a small proportion of ordinary matter accreted within
the mirror matter SB).
Furthermore, the internal heating of the
mirror SB by the interactions of the atmospheric atoms
within the SB may actually cause the required atmospheric explosion.

It is also interesting to note that there is evidence
that smaller `Tunguska-like' events are actually quite common,
occurring on a yearly basis. Such events have
been catalogued by Ol'khovatov\cite{andr} with the
most recent such event occurring only a few months
ago in Jordan\cite{jordan}. 
There are many events (see e.g. Ref.\cite{docobo,jordan})
where low altitude `fireballs'
are observed, yet such fireballs (if they are
due to an ordinary matter SB) should originate
from huge and enormously bright fireballs higher up in 
the atmosphere because of ablation and fragmentation. 
These bright parents of low altitude fireballs are inexplicably not
observed. Even more remarkable is that these
`fireballs' have been observed in some cases to actually hit the
ground (we will discuss an explicit example of this
in a moment),  yet no meteorite fragments 
were recovered.
The strange properties of these events has
lead to purely geophysical explanations. For example,
it has been proposed that they are due to
some poorly understood coupling between tectonic 
and atmospheric process rather than to
some type of SB\cite{andr}.
Mirror matter represents an exciting and fun
alternative possibility which can be tested in a number
of ways as we will now briefly discuss.

First, it requires large photon - mirror 
photon kinetic mixing of the order given
by the orthopositronium experiment for the
mirror SB to release its energy in
the atmosphere. Thus, we
could simply repeat the orthopositronium experiment
to make sure that mirror matter exists with
the required kinetic mixing. 
More work could be done in trying to understand the
detailed properties of mirror matter space bodies
interacting with the Earth's atmosphere which
might allow the idea to be more rigorously compared
with observations.  For example, the 1997 Greenland event was
observed with satellites and a ground based video camera\cite{ped}.
This event has been estimated
to be due to a 36,000 Kg SB which
fragmented and exploded over Greenland. 
No fragments or even meteoritic dust in the snow was
found by search teams\cite{ped}.
The study\cite{ped} also found that
the SB had an anomalous ablation coefficient\cite{ped}
which might be something which could be used
to possibly test the mirror matter hypothesis for these
space bodies.  

Perhaps the most spectacular
way to test the idea though is to actually find it!
Mirror matter could be searched for
in the ground at the various impact sites.
Any mirror matter fragments would have melted when they
hit the ground
and reformed becoming mixed with ordinary matter at
some distance underground. 
The small effective ordinary electric charges of the mirror
electrons ($\epsilon e$) which is given to them
by the photon - mirror photon kinetic mixing should
easily lead to enough electrostatic repulsion 
(which is linear in $\epsilon$) to resist gravity
which means that the mirror matter will eventually stop
(if it solidifies).  There may be some amount close
to the surface which could potentially be extracted and 
purified.  Importantly, many of these sites are very localized 
and very accessible. For example, in the recent Tunguska-like
event which occurred in Jordan (about 50 kilometers from
the capital Amman) only a few months ago
\cite{jordan} the fireball was observed
(by a crowd of about 100 people in a funeral procession) 
to break up into two pieces and observed to actually hit the ground! 
The two sites where the `objects'
landed featured a half burnt tree and
a half burnt rock (see Ref.\cite{jordan} for the 
remarkable pictures) but no ordinary crater and no ordinary matter fragments.
\footnote{
Potentially a mirror matter SB could leave a type of
impact crater depending on the chemical composition of
the SB and also on the nature of the Earth's surface at
the impact site.} 
One could take samples of earth below the burnt tree (or the parts
of the burnt tree itself) and try to extract mirror
atoms. This might be possible by taking samples
and putting them into a centrifuge which should allow
the mirror matter to be separated from the ordinary
matter (or at least greatly purified).
It would be a very exciting experiment and lots of
fun too!

Finally, mirror matter should have all sorts of useful 
industrial applications.  Of course it is premature to 
speculate too much along these lines until it is actually 
discovered, but the point is that its possible existence is 
not merely of interest to people who want to understand the 
fundamental laws of nature or find out
what the Universe is made of.
Unlike Higgs particles or top quarks it may actually 
be a very useful new material with all sorts
of practical applications. This provides another important 
motivation to search for it, either by repeating the orthopositronium 
experiment in vacuum or by digging it out of the ground.
Of course I love Higgs particles and top quarks too but
it is also important to remember that
pure research in particle and astrophysics can sometimes 
lead to discoveries with widespread implications for
society, in addition to the intrinsic merits and long
term importance of such pure science itself.


\vskip 0.4cm
\noindent
{\bf Acknowledgement}
\vskip 0.4cm
\noindent
The author is an Australian Research Fellow.
The author would like to thank Z. Ceplecha for
patiently answering some of my questions.

\end{document}